\documentclass[12pt]{article}

\usepackage{amsfonts}
\usepackage{amsmath}
\usepackage{epsfig}
\usepackage{latexsym}
\usepackage{graphicx}
\usepackage{subfigure}
\usepackage{amssymb}
\numberwithin{equation}{section}
\allowdisplaybreaks

\def\spa#1{\phantom{\fbox{\rule[-#1cm]{0cm}{0cm}}}}

\def\be{\begin{equation}}
\def\ee{\end{equation}}
\def\bea{\begin{eqnarray}}
\def\eea{\end{eqnarray}}
\def\bequ{\begin{equation}}
\def\eequ{\end{equation}}

\def\del{\partial}

\renewcommand{\thefootnote}{\fnsymbol{footnote}}

\begin{document}

\hfuzz=100pt
\title{
\begin{flushright} 
\small{KEK-TH-1644}
\end{flushright} 
{\Large \bf{PPN expansion and FRW scalar perturbations \\  in $n$-DBI gravity}}}
\date{}
\author{Fl\'avio S. Coelho$^a$, Carlos Herdeiro$^a$, Shinji Hirano$^{b}$ and Yuki Sato$^{c} \ $\footnote{Emails: flavio@physics.org, herdeiro@ua.pt, shinji.hirano@wits.ac.za, satoyuki@post.kek.jp}
  \spa{0.5} \\
\\
$^a${\small{\it Departamento de F\'isica da Universidade de Aveiro and I3N}}
\\ {\small{\it Campus de Santiago, 3810-183 Aveiro, Portugal}}\\
$^{b}${\small{\it School of Physics and Center for Theoretical Physics, University of the Witwatersrand}}
\\ {\small{\it WITS 2050, Johannesburg, South Africa}}
\\ 
$^{c}${\small{\it KEK Theory Center, High Energy Accelerator Research Organization (KEK)  } }
\\ {\small{ \it  Oho-ho, Tsukuba, Ibaraki 305-0801, Japan  }  }
}

\maketitle
\centerline{}

\begin{abstract}
$n$-DBI gravity explicitly breaks Lorentz invariance by the introduction of a unit time-like vector field, thereby giving rise to an extra (scalar) degree of freedom. We look for observational consequences of this mode in two setups. Firstly, we compute the parametrized post-Newtonian (PPN) expansion of the metric to first post-Newtonian order. Surprisingly, we find that the PPN parameters are exactly the same as in General Relativity (GR), and no preferred-frame effects are produced. In particular this means that $n$-DBI gravity is consistent with all GR solar system experimental tests. We discuss the origin of such degeneracy between $n$-DBI gravity and GR, and suggest it may also hold in higher post-Newtonian order. Secondly, we study gravitational scalar perturbations of a Friedmann-Robertson-Walker space-time with a cosmological constant $\Lambda\geq0$. 
In the case of de Sitter space, we show that the scalar mode grows as the universe expands and, in contrast with a canonical scalar field coupled to GR, it does not freeze on superhorizon scales. 

\end{abstract}

\renewcommand{\thefootnote}{\arabic{footnote}}
\setcounter{footnote}{0}

\newpage
\tableofcontents

\section{Introduction}
\label{introduction}

After an almost century-old quest, General Relativity (GR) continues to stand as the best description of gravitational phenomena. Numerous attempts to generalize it have been made, motivated either by the development of observational and precision cosmology, by the desire to make it compatible with quantum mechanics or simply to explore its theoretical properties and better understand how special and unique it is. 

Recently, there has been an increased interest in theories which break the gauge symmetry of GR (full invariance under coordinate transformations or diffeomorphisms) down to the sub-group of foliation-preserving diffeomorphisms, mostly motivated by Ho\v{r}ava's proposal \cite{Horava:2009uw} of an anisotropic scaling of space and time as an attempt to produce an ultraviolet (UV) completion of GR, whilst retaining its properties in the infrared (IR).

Any explicit breaking of general diffeomorphism invariance forcefully gives rise to extra degrees of freedom in a theory of gravity. This violation of Lorentz symmetry is no exception and induces a scalar mode describing excitations in the foliation structure of space-time.

Another theory in this class, $n$-DBI gravity \cite{Herdeiro:2011km}-\cite{Coelho:2013aa}, was motivated by scale invariance and designed to reduce to the Dirac-Born-Infeld (DBI) scalar field theory for homogeneous conformally flat geometries. Having only two dimensionless parameters, $\lambda$ and $q$, it nicely accommodates two accelerated epochs mediated by radiation- and matter-dominated periods, with a natural hierarchy between the two effective cosmological constants \cite{Herdeiro:2011km}. The dynamics of its scalar mode were extensively studied in \cite{Coelho:2012xi}, where its existence as a full degree of freedom was established and none of the pathologies associated with similar models were found. Its physical action, however, remains somewhat elusive.

The purpose of this paper is to explore the physical action and possible observational consequences of this scalar graviton in $n$-DBI gravity, and we shall pursue this goal using two different directions. Firstly we will consider the parametrized post-Newtonian (PPN) framework applied to $n$-DBI gravity. Secondly we shall study cosmological perturbations of $n$-DBI gravity around Friedmann-Robertson-Walker (FRW) models.

The PPN formalism (see \cite{Will:2005va} for an overview) is a general framework which allows for a systematic comparison of different theories of gravity in the weak field, slow velocities regime. Each theory is then fully characterized by ten dimensionless parameters, each of which can be ascribed a distinct meaning or effect. Of particular interest to us are the parameters $\alpha_1$ and $\alpha_2$, known to be responsible for preferred-frame effects.\footnote{The parameter $\alpha_3$ belongs to the same category but is necessarily absent in semi-conservative theories, derived from a covariant Lagrangian.}  They vanish in GR but are expected to be non-zero in theories with a preferred foliation. Indeed, they were computed in the context of the IR limit of Ho\v{r}ava-Lifshitz theory \cite{Blas:2010hb} (also known as khronometric theory) and Einstein-aether theory \cite{Foster:2005dk}. Given all the free parameters of these theories, however, it is not very surprising that constraints can be imposed such that the experimental bounds on $\alpha_1$ and $\alpha_2$ are obeyed (see \cite{Shao:2012xc}-\cite{Iorio:2012pe} for recent numbers).

$n$-DBI gravity presents a potentially more interesting scenario since it has only two free parameters. In \cite{Herdeiro:2011km} the parameters $\lambda$ and $q$ were estimated to be
\begin{equation}
\lambda_{inf}\sim10^{-8}\,,\qquad q\sim1+10^{-110}\,,
\label{phenovalues}
\end{equation}
in order to produce the required orders of magnitude for the scale of inflation and for the late-time cosmological constant. In particular one observes that the dimensionless parameter $\lambda$ is small. On the other hand, GR is recovered in the limit $\lambda\rightarrow\infty$ \cite{Herdeiro:2011im}. Since GR is a good theory in the weak field limit we expect a lower bound for $\lambda$ to be provided by a post-Newtonian analysis. This makes the PPN analysis for $n$-DBI gravity particularly interesting, since a constraint incompatible with the value required by cosmology would rule out the theory whereas a constraint compatible with that value would increase the appeal of the theory.  

The result of our investigation is, in a sense, neither of these two possibilities, which is rather surprising. We find that there is no constraint on $\lambda,q$. The central point is that there is a subclass of solutions of the theory that coincides with those of GR and that exists for any $\lambda,q$; these parameters are therefore left unconstrained. This was already observed at the level of black hole solutions \cite{Herdeiro:2011im,Coelho:2013zq}. Herein we observe this subclass of solutions is large enough to include all of GR at PN level. Hence, $n$-DBI gravity looks indistinguishable from GR, at least at first post-Newtonian order, and in particular there are no preferred-frame effects. In this analysis we conclude the scalar graviton eludes us again.

In the second part of this paper, we look for observational consequences of the scalar graviton in a different setup: we study scalar perturbations in a time-dependent, spatially homogeneous and isotropic cosmological model. We start by reviewing the flat space-time case \cite{Coelho:2012xi}
and then specialize the FRW model to represent a patch of de Sitter space-time. We study the scalar mode and show
that it endows the metric with a spatially-arbitrary perturbation which grows in time as the universe expands.
This is a novel feature, suggesting that, in contrast to a scalar field propagating in a de Sitter universe in GR, the 
perturbations of the scalar mode in $n$-DBI gravity do not freeze on super-horizon scales. This fact may find its origin in the lack of Lorentz invariance in this model, and introduces a clear distinction with GR when considering the concept and computation of the density perturbations spectrum sourced by the scalar mode perturbations. 

This paper is organized as follows. A short review of the basic equations of $n$-DBI gravity, in one of its formulations  \cite{Coelho:2012xi}, is provided in Sec. \ref{intro_ppn}. The PPN analysis is performed in Sec. \ref{ppn}, wherein a self-contained review of the method is provided, which turns out to be applicable to the case under study, \textit{mutatis mutandis}. Sec. \ref{scalarfrw} is devoted to the analysis of gravitational scalar perturbations in $n$-DBI gravity.
Concluding remarks and a discussion are presented in Sec. \ref{conclusion}.

\section{Basic equations for $n$-DBI gravity}
\label{intro_ppn}
In the linearized form of \cite{Coelho:2012xi}, $n$-DBI gravity is defined by the action
\begin{equation}
S=-\frac{1}{16\pi G_N}\int d^4x\sqrt{-g}\, e\left(\mathcal{R}-2G_N\Lambda(e)\right)+S_{matter}\,,
\label{action}
\end{equation}
where we use natural units in which $c=1$, $G_N$ is Newton's constant,
\begin{eqnarray}
\mathcal{R}&=&R-2\nabla_\mu(n^\mu \nabla_\nu n^\nu)\,,\label{R_def}\\
\Lambda(e)&=&\frac{3\lambda}{G_N^2}\left(\frac{2q}{e}-1-\frac{1}{e^2}\right)\,,
\end{eqnarray}
$n^\mu$ is the everywhere normalized time-like vector field defining the space-time foliation, $\lambda,q$ are the two parameters of $n$-DBI gravity \cite{Herdeiro:2011km} and $e$ is an auxiliary field whose equation of motion reads
\begin{equation}
e=\left(1+\frac{G_N}{6\lambda}\mathcal{R}\right)^{-1/2}\,.\label{e}
\end{equation}

We will also work in the covariant formulation of  $n$-DBI gravity \cite{Coelho:2012xi} by defining the vector field $n^\mu$ through a St\"uckelberg field $\varphi$, dubbed \textit{khronon field}
\begin{equation}
n_\mu=-\frac{\del_\mu\varphi}{\sqrt{-X}}\,,\qquad X\equiv g^{\mu\nu}\del_\mu\varphi\del_\nu\varphi\,.
\end{equation}

For convenience, we denote 
\begin{equation}
K_{\mu\nu}\equiv \nabla_\mu n_\nu\,, \qquad a_\mu\equiv n^\nu\nabla_\nu n_\mu\, ,
\end{equation}
where the latter quantity is the acceleration of $n$, and the trace of $K_{\mu\nu}$ by $K$.

Variation of (\ref{action}) with respect to the metric yields the field equations
\begin{equation}
eR_{\mu\nu}=g_{\mu\nu}G_Ne\Lambda(e)+\left(T^e_{\mu\nu}-\frac{1}{2}g_{\mu\nu}T^e\right)+\left(T^\varphi_{\mu\nu}-\frac{1}{2}g_{\mu\nu}T^\varphi\right)+8\pi G_N\left(T_{\mu\nu}^{mat}-\frac{1}{2}g_{\mu\nu}T^{mat}\right)\,,\label{field_eq}
\end{equation}
where
\begin{eqnarray}
T^e_{\mu\nu}&=&\nabla_\mu\del_\nu e-g_{\mu\nu}\nabla^\sigma\del_\sigma e\,,\label{T_e}\\
T^\varphi_{\mu\nu}&=&-g_{\mu\nu}\left(a^\sigma\del_\sigma e+n^\sigma n^\rho\nabla_\sigma\del_\rho e\right)+n_\mu n_\nu\left(n^\sigma n^\rho\nabla_\sigma\del_\rho e-K n^\sigma\del_\sigma e+a^\sigma\del_\sigma e\right)\nonumber\\
&&+2n_{(\mu}\left(K_{\nu)\sigma}\del^{\sigma}e+ n^\sigma \nabla_{\nu)}\del_\sigma e- K \del_{\nu)}e\right)\,.\label{T_phi}
\end{eqnarray}

The equation for the khronon field $\varphi$ can be written as the conservation of a current, 
\bequ
\nabla_\mu J^\mu_\varphi=0 \ , 
\label{khronon_eq}
\eequ
where
\begin{equation}
\sqrt{-X}J^\mu_\varphi=2(g^{\mu\nu}+n^\mu n^\nu)\left(K_{\nu\sigma}\del^\sigma e-K\del_\nu e+n^\sigma\nabla_\sigma\del_\nu e\right)\,.
\end{equation}
This follows from the conservation of the energy-momentum tensor for the khronon field
and is also equivalent to the extra equation of  \cite{Coelho:2012xi} obtained through the time-derivative of the Hamiltonian constraint which, unlike in GR, is not automatically preserved by time-evolution in the $3+1$ formulation of $n$-DBI gravity.

\section{PPN framework for $n$-DBI gravity}
\label{ppn}
$n$-DBI gravity is a metric theory of gravity (see \cite{Will:2005va} for a discussion of the postulates obeyed by metric theories of gravity). As such, the post-Newtonian effects for $n$-DBI, that are tested by solar system observations, can be discussed using the parameterized post-Newtonian (PPN) framework. In this section we discuss the PPN framework and its application to $n$-DBI gravity.

\subsection{Summary of the PPN framework}
The PPN framework (see \cite{Will:2005va} for an overview) is a perturbative and iterative scheme to provide a solution of the field equations of a given metric theory of gravity, where it is assumed there exists a suitable small parameter $\varepsilon$, which reflects both a slow-motion and a weak field limit of the theory. A quantity $X$ is then said to be of order $n$, denoted $\mathcal{O}(n)$, if it is of order $n$ in $\varepsilon$, and the expansion in $\varepsilon$ is applied to both sides of the field equations.

Concerning the right hand side, matter is usually taken to be a perfect fluid with energy-momentum
\begin{equation}
T_{\mu\nu}^{mat}=(\rho(1+\Pi)+p)u_\mu u_\nu+p g_{\mu\nu}\,,
\label{tmat}
\end{equation}
where $u^\mu=dx^\mu/d\tau$ is the $4$-velocity of an element of fluid, with 3-velocity $v^i$, $\rho$ is the rest mass density of the element, $\rho\Pi$ is its internal energy density ans $p$ its total pressure. The following dimensionless quantities are then assumed to have the orders:
\begin{equation}
U\sim v^2\sim p/\rho\sim\Pi\sim \mathcal{O}(2)\,,\qquad \left|\frac{\partial / \partial t}{\partial / \partial x}\right|\sim \mathcal{O}(1) \ ,
\label{basicorders}
\end{equation}
where $U$ is (minus) the Newtonian gravitational potential. In the solar system $\varepsilon^2 \lesssim 10^{-5}$ and thus the perturbative expansion is justified. Observe that time derivatives effectively increase the order of smallness.

Concerning the left hand side, the metric is expanded around the background cosmological solution (Minkowski space-time is a good approximation at the solar system level); the same applies to any additional fields present in the theory. Requiring a matching with Newtonian gravity, i.e. considering the Newtonian limit for the given metric theory of gravity including consistency with its conservation laws, determines the lowest order terms for the metric coefficients:
\bequ
g_{00}=-1+2U+ \mathcal{O}(4) \ , \qquad g_{0i}=0+ \mathcal{O}(3) \ , \qquad g_{ij}=\delta_{ij}+\mathcal{O}(2) \ .
\label{newlim}
\eequ
Thus, the post-Newtonian corrections appear at order $\mathcal{O}(4)$ for  $g_{00}$,  $\mathcal{O}(3)$ for $g_{0i}$ and $\mathcal{O}(2)$ for $g_{ij}$. The goal of the PPN framework is to compute these corrections for the metric theory of gravity under study. To do so, a concrete form of the metric is constructed, to facilitate the comparison between different metric theories of gravity as well as comparison with experiment. Such form of the metric is obtained in three steps. First, a set of \textit{post-Newtonian potentials} is introduced. These potentials are denoted:
\[ \Phi_W, \Phi_1,\Phi_2,\Phi_3,\Phi_4,\mathcal{A},\mathcal{B} \sim \mathcal{O}(4); \ \ \ \  V_i,W_i \sim \mathcal{O}(3); \ \ \ \  U_{ij}\sim \mathcal{O}(2);
\]
where the potentials have been separated according to being scalars, vectors and tensors, respectively, under the rotation group. These potentials encode the functional dependence that can arise, to this order, in the metric due to the matter sources, leaving as the only remaining freedom the coefficients with which the potentials enter the metric.\footnote{To give a concrete example (the definitions of all post-Newtonian potentials can be found in \cite{Will:2005va})
\[ \Phi_1\equiv \int \frac{\rho'v'^2}{|{\bf x}-{\bf x}'|}d^3x' \ .
\] It should be understood that there is an infinite number of such possible potentials and thus the chosen expansion, albeit justified by reasonable arguments, is really an \textit{ansatz}.} 
Second, a convenient coordinate system is chosen, called \textit{standard post-Newtonian gauge}, in which the spatial part of the metric is isotropic, thus eliminating the potential $U_{ij}$, and in which the potential $\mathcal{B}$ is gauged away from $g_{00}$. The PPN metric coefficients then include ten post-Newtonian corrections and thus ten coefficients are needed:
\begin{eqnarray}
g_{00}&=&-1+2U-2\beta U^2-2\xi\Phi_W+(2\gamma+2+\alpha_3+\zeta_1-2\xi)\Phi_1 \nonumber\\
&&+2(3\gamma-2\beta+1+\zeta_2+\xi)\Phi_2+2(1+\zeta_3)\Phi_3+2(3\gamma+3\zeta_4-2\xi)\Phi_4-(\zeta_1-2\xi)\mathcal{A} \nonumber \\
&&+\mathcal{O}(5)\,, \label{ppn1} \\
g_{0i}&=&-\frac{1}{2}(4\gamma+3+\alpha_1-\alpha_2+\zeta_1-2\xi)V_i-\frac{1}{2}(1+\alpha_2-\zeta_1+2\xi)W_i+\mathcal{O}(4)\,,\\
g_{ij}&=&(1+2\gamma U)\delta_{ij}+\mathcal{O}(3) \label{ppn3}\,.
\label{ppnpot}
\end{eqnarray}
The ten constants, called the \textit{PPN parameters},
\[
\alpha_1,\alpha_2,\alpha_3, \beta, \gamma,\xi, \zeta_1,\zeta_2,\zeta_3,\zeta_4 \ ,
\]
enter the metric in various linear combinations, designed to provide a cleaner intepretation for each of the PPN parameters, that we shall discuss shortly. The numerical value of these parameters can then be established for any theory and compared with experiments. GR, for instance, has $\beta=\gamma=1$ and all others vanish.

Observe that the next order correction in \eqref{ppn1}-\eqref{ppn3} beyond the post-Newtonian terms - dubbed \textit{the post-post Newtonian terms} -  are of order  $\mathcal{O}(5)$ for  $g_{00}$,  $\mathcal{O}(4)$ for $g_{0i}$ and $\mathcal{O}(3)$ for $g_{ij}$; in particular cases, for instance if post-Newtonian energy is conserved, the next non-trivial term will actually be one order higher: $\mathcal{O}(6)$ for  $g_{00}$,  $\mathcal{O}(5)$ for $g_{0i}$ and $\mathcal{O}(4)$ for $g_{ij}$. The natural dissipative effect of relativistic gravity, i.e. gravitational radiation emission, occurs only at higher order in this expansion.

The metric form \eqref{ppn1}-\eqref{ppn3} uses a quasi-Cartesian coordinate system $(t,{\bf x})$, whose outer regions are at rest with respect to the Universe rest frame, i.e.  a frame in which the Universe appears isotropic. The third and final step is to change to a moving coordinate system, with velocity ${\bf w}\sim \mathcal{O}(1)$, with respect to the Universe rest frame. This is achieved by performing a Lorentz transformation, to the appropriate order. The final result for the PPN metric can still be expressed in standard post-Newtonian gauge, where now all quantities refer to the new coordinates: 
\begin{eqnarray}
g_{00}&=&-1+2U-2\beta U^2-2\xi\Phi_W+(2\gamma+2+\alpha_3+\zeta_1-2\xi)\Phi_1 \nonumber \\
&&+2(3\gamma-2\beta+1+\zeta_2+\xi)\Phi_2+2(1+\zeta_3)\Phi_3+2(3\gamma+3\zeta_4-2\xi)\Phi_4-(\zeta_1-2\xi)\mathcal{A}\nonumber \\
&&-(\alpha_1-\alpha_2-\alpha_3)w^2U-\alpha_2w^iw^jU_{ij}+(2\alpha_3-\alpha_1)w^iV_i+\mathcal{O}(5)\,, \label{ppn12}\\
g_{0i}&=&-\frac{1}{2}(4\gamma+3+\alpha_1-\alpha_2+\zeta_1-2\xi)V_i-\frac{1}{2}(1+\alpha_2-\zeta_1+2\xi)W_i \nonumber \\
&&-\frac{1}{2}(\alpha_1-2\alpha_2)w^iU-\alpha_2w^jU_{ij}+\mathcal{O}(4)\,,\\
g_{ij}&=&(1+2\gamma U)\delta_{ij}+\mathcal{O}(3) \label{ppn32}\,.
\label{ppnpot2}
\end{eqnarray}

A first observation to interpret the PPN parameters can be obtained by inspection of   \eqref{ppn12}-\eqref{ppn32}: if any of $\alpha_1,\alpha_2,\alpha_3$ are non-zero, there are observable effects that depend on the velocity ${\bf w}$ with respect to the preferred frame of the Universe. Thus, the PPN parameters  $\alpha_1,\alpha_2,\alpha_3$ are interpreted as measuring preferred frame effects. 

An interpretation of the PPN parameters $\alpha_3,\zeta_1,\zeta_2,\zeta_3,\zeta_4$ is given by considering energy momentum conservation at post-Newtonian level: they measure the violation of total energy and momentum in a metric theory of gravity. It was shown that these parameters vanish for metric theories of gravity derived from an action principle.

Theories with vanishing $\alpha_1,\alpha_2,\alpha_3,\zeta_1,\zeta_2,\zeta_3,\zeta_4$ conserve \textit{also} total angular momentum and are called \textit{fully conservative}. Such theories have only three non-trivial PPN parameters: $\gamma$, which measures the space-curvature; $\beta$, which measures the non-linearity of the theory to this order; and $\xi$, which measures the existence of preferred location effects. Note that these interpretations are not covariant (except for $\gamma$) and hold only in the standard post-Newtonian gauge.

In $n$-DBI , $\alpha_3,\zeta_1,\zeta_2,\zeta_3,\zeta_4$ all vanish, since it is a Lagrangian based theory, defined by \eqref{action}. But naively, one may expect that due to the breakdown of local Lorentz invariance associated to the existence of the vector field $n^\mu$, a non-zero $\alpha_1$ and/or $\alpha_2$ may exist, since these parameters measure preferred-frame effects. We shall see in the next section, however, that this naive expectation is not confirmed.

\subsection{PPN parameters for $n$-DBI gravity}
The computation of the PPN parameters for a given metric theory of gravity follows a well defined recipe. We shall now describe the method, which again has three steps, using the case of GR for concreteness. As we shall see shortly, the case of $n$-DBI will reduce to this one.

The first step is to identify the relevant variables and expand them to post Newtonian order. In the case of GR these are solely the metric $g_{\mu\nu}$ and the matter, described by the energy momentum tensor \eqref{tmat}. 
We take the background to be Minkowski space-time with metric $\eta_{\mu\nu}$, in a global, almost inertial chart $x^\mu=(t,x^i)$. The PPN expansion for the metric is  
\begin{equation}
g_{\mu\nu}=\eta_{\mu\nu}+h_{\mu\nu} \ ,
\label{ppn_g1}
\end{equation}
where
\bequ
h_{00}=h_{00}^{(2)}+h_{00}^{(4)} \ , \qquad h_{0i}=h_{0i}^{(3)} \ ,\qquad h_{ij}=h_{ij}^{(2)} \ .
\label{ppn_g2}
\eequ
Observe that we have denoted the post-Newtonian order of each term by a superscript. The matter energy momentum tensor, on the other hand, is not dimensionless. But extracting the mass density one obtains a dimensionless quantity
\begin{equation}
T_{\mu\nu}^{mat}=\rho\left[(1+\Pi+\frac{p}{\rho})u_\mu u_\nu+\frac{p}{\rho} g_{\mu\nu}\right]\, ;
\label{tmat2}
\end{equation}
the expansion of each component then yields
\bequ
T_{00}^{mat}=\rho\left[1+\Pi-h_{00}^{(2)}+v^2+\mathcal{O}(4)\right]\ , 
\label{tmat21}
\eequ
\bequ
 T_{0i}^{mat}=-\rho\left[v_i+\mathcal{O}(3)\right] \ , \qquad T_{ij}^{mat}=\rho\left[v_iv_j+\frac{p}{\rho}\delta_{ij}+\mathcal{O}(4)\right] \ .
 \label{tmat22}
\eequ
It follows from these components and \eqref{ppn_g1}-\eqref{ppn_g2} that the trace of the matter energy-momentum tensor is (which actually coincides with the exact result)
\bequ
T^{mat}=\rho\left[-1-\Pi+\frac{3p}{\rho}+\mathcal{O}(4)\right]\ . 
\eequ
The orders given are sufficient for the computation of the PPN parameters.

The second step is to substitute \eqref{ppn_g1}-\eqref{ppn_g2} and \eqref{tmat21}-\eqref{tmat22} in the field equations, 
\bequ
R_{\mu\nu}=8\pi G_N\left(T_{\mu\nu}^{mat}-\frac{1}{2}g_{\mu\nu}T^{mat}\right)\,,\label{field_eq_gr} 
\eequ
keeping the necessary terms to obtain a consistent PPN solution. This turns out to require including the following terms for the various components of the Ricci tensor:
\begin{eqnarray}
R_{00}&=&-\frac{1}{2}h_{00,ii}-\frac{1}{2}(h_{jj,00}-2h_{j0,j0})+\frac{1}{2}h_{jk}h_{00,jk}-\frac{1}{4}h_{00,i}h_{00,i}+\frac{1}{2}h_{00,j}(h_{jk,k}-\frac{1}{2}h_{kk,j})\,, \nonumber\\
R_{0j}&=&-\frac{1}{2}(h_{0j,ii}-h_{k0,jk}+h_{kk,0j}-h_{kj,0k})\,,\nonumber \\
R_{ij}&=&-\frac{1}{2}(h_{ij,kk}-h_{00,ij}+h_{kk,ij}-h_{ki,kj}-h_{jk,ki})\,.
\end{eqnarray}
Observe that there are both linear and quadratic terms in the metric perturbation around Minkowski space-time \eqref{ppn_g2}. The criterion for neglecting other terms take both into account the orders in \eqref{ppn_g2} and in the derivatives \eqref{basicorders}. Then, separating different orders, the field equations \eqref{field_eq_gr}  yield
\bequ
\Delta h_{00}^{(2)}=-8\pi G_N\rho \ ,
\label{ppngr1}
\eequ
and
\begin{eqnarray}
&&-\Delta h_{00}^{(4)}-h_{jj,00}^{(2)}+2h_{j0,j0}^{(3)}+h_{jk}^{(2)}h_{00,jk}^{(2)}+h_{00,j}^{(2)}\left(-\frac{1}{2}h_{00,j}^{(2)}+h_{jk,k}^{(2)}-\frac{1}{2}h_{kk,j}^{(2)}\right) \nonumber \\
&&= 8\pi G_N\rho \left[\Pi-h_{00}^{(2)}+2v^2+\frac{3p}{\rho}\right] \ ,
\label{ppngr2}
\end{eqnarray}
both from the $00$ component. We have denoted the Laplacian on $\mathbb{R}^3$ by $\Delta$. From the $0j$ component we get
\bequ
\Delta h_{0j}^{(3)}-h_{k0,jk}^{(3)}+h_{kk,0j}^{(2)}-h_{kj,0k}^{(2)}=16\pi G_N\rho v_j \ ,
\label{ppngr3}
\eequ
and finally from the $ij$ component we find
\bequ
\Delta h_{ij}^{(2)}-h_{00,ij}^{(2)}+h_{kk,ij}^{(2)}-h_{ki,kj}^{(2)}-h_{jk,ki}^{(2)}=-8 \pi G_N \rho \delta_{ij} \ .
\label{ppngr4}
\eequ

The third step is to solve the resulting equations \eqref{ppngr1}-\eqref{ppngr4}, using a specific order that starts from lower to higher order equations and making appropriate gauge choices. We start by solving the $00$ equation to $\mathcal{O}(2)$, i.e. \eqref{ppngr1}, which by comparison with Poisson's equation in Newtonian gravity yields
\bequ
h_{00}^{(2)}=2U \ , 
\label{newper}
\eequ
in agreement with \eqref{newlim}. This is the Newtonian limit and yields no information about the PPN parameters. Next we solve the $ij$ equation to $\mathcal{O}(2)$. This becomes very simple if we choose three gauge conditions corresponding to the spacial part of the de Donder gauge: $h^\mu_{\ k,\mu}-\frac{1}{2}\left(h_{\mu\alpha}\eta^{\mu\alpha}\right)_{,k}=0$. This gauge condition, to $\mathcal{O}(2)$, simplifies \eqref{ppngr4} to
\bequ
\Delta h_{ij}^{(2)}=-8 \pi G_N \rho \delta_{ij} \ \ \ \  \Rightarrow \ \ \ \   h_{ij}^{(2)}=2U\delta_{ij} \ .
\label{ppngr42}
\eequ
Comparison with \eqref{ppnpot2} determines the first PPN parameter: $\gamma=1$. Next we solve the $0j$ equation to $\mathcal{O}(3)$. We use the gauge condition $h^\mu_{\ 0,\mu}-\frac{1}{2}\left(h_{\mu\alpha}\eta^{\mu\alpha}\right)_{,0}=-\frac{1}{2}h_{00,0}$, which is \textit{not} the temporal component of the de Donder gauge condition. Making use of the solutions \eqref{newper} and \eqref{ppngr42} the equation \eqref{ppngr3} becomes
\bequ
\Delta h_{0j}^{(3)}+U_{,0j}=16\pi G_N\rho v_j  \ \ \ \  \Rightarrow \ \ \ \   h_{0j}^{(3)}=-\frac{7}{2}V_j-\frac{1}{2}W_j \ ,
\label{ppngr32}
\eequ
since the PPN potentials $V_j$ and $W_j$ obey
\bequ
\Delta V_j=-4\pi G_N\rho v_j \ , \qquad \Delta(V_j-W_j)=-2U_{,0j} \ .
\eequ
It can be checked that this solution obeys the imposed gauge condition as it should. Since GR is derived from an action we know that $\alpha_3,\zeta_1,\zeta_2,\zeta_3,\zeta_4=0$; then comparison of \eqref{ppngr32} with \eqref{ppnpot2} yields $\alpha_1=\alpha_2=\xi=0$. Finally we solve the $00$ equation to $\mathcal{O}(4)$. Using the lower order solutions and gauge conditions \eqref{ppngr2} becomes
\bequ
-\Delta\left(h_{00}^{(4)}+2U^2\right)= 8\pi G_N\rho \left[\Pi+2U+2v^2+\frac{3p}{\rho}\right]   \ \ \ \  \Rightarrow \ \ \ \  h_{00}^{(4)}=-2U^2+4\Phi_1+4\Phi_2+2\Phi_3+6\Phi_4 \ , 
\eequ
since the PPN potentials $\Phi_1,\Phi_2,\Phi_3,\Phi_4$ obey
\bequ
\Delta \Phi_1=-4\pi\rho v^2 \ , \qquad \Delta \Phi_2=-4\pi\rho U \ , \qquad \Delta \Phi_3=-4\pi\rho \Pi \ , \qquad \Delta \Phi_4=-4\pi p \ .
\eequ
Comparison with \eqref{ppnpot2} determines the final PPN parameter: $\beta=1$.

We now turn our attention to $n$-DBI. The relevant variables are the same as in GR plus the khronon field $\varphi$. The background value of the khronon is $\varphi=t$, where $t$ is the time coordinate in the standard PPN gauge. Thus, considering a first order perturbation to this value we have
\begin{equation}
\varphi=t+\chi^{(1)}\, . 
\end{equation}
The expansions of some other quantities relevant for computing the field equations  \eqref{field_eq} and \eqref{khronon_eq} to PPN order are:
\begin{eqnarray}
\frac{1}{\sqrt{-X}}&=&1-\dot{\chi}^{(1)}-\frac{1}{2}h_{00}^{(2)}+\frac{1}2{}\del_i\chi^{(1)}\del_i\chi^{(1)}\,,\\
n_0&=&-1+\frac{1}{2}h^{(2)}_{00}-\frac{1}{2}\del_i\chi^{(1)}\del_i\chi^{(1)} \,, \qquad n_i=-\del_i\chi^{(1)} \,,\\
n^0&=&1+\frac{1}{2}h^{(2)}_{00}+\frac{1}{2}\del_i\chi^{(1)}\del_i\chi^{(1)} \, , \qquad
n^i=-\del_i\chi^{(1)}\,,\\
K_{00}&=&0\,, \qquad
K_{0i}=-\del_i\left(\dot{\chi}^{(1)}+\frac{1}{2}h_{00}^{(2)}\right)\,, \\
K_{i0}&=&-\frac{1}{2}\del_i\left(\del_j\chi^{(1)}\del_j\chi^{(1)}\right)\,,\qquad
K_{ij}=-\del_i\del_j\chi^{(1)}\,,\\
K&=&-\Delta\chi^{(1)} \,,\\
a_0&=&0\,,\qquad
a_i=-\del_i\left(\dot{\chi}^{(1)}+\frac{1}{2}h_{00}^{(2)}-\frac{1}{2}\del_j\chi^{(1)}\del_j\chi^{(1)}\right)\,,\\
\nabla_\mu a^\mu&=&-\Delta\left(\dot{\chi}^{(1)}+\frac{1}{2}h_{00}^{(2)}-\frac{1}{2}\del_j\chi^{(1)}\del_j\chi^{(1)}\right) \,,\label{cr1}\\
\mathcal{R}&=&\Delta h_{00}^{(2)}-\delta_{ij}\Delta h_{ij}^{(2)}+\partial_{ij}h_{ij}^{(2)}-2\Delta\chi^{(1)}\Delta\chi^{(1)}+2\Delta\dot{\chi}^{(1)}-2\partial_i\chi^{(1)}\partial_i\Delta \chi^{(1)} \, ,\label{cr2} \\
J_0^\varphi&=&0 \, , \qquad J_i^\varphi=2\left[ \partial_i\dot{e}^{(2)} +\partial_ie^{(2)}\Delta\chi^{(1)}-\partial_i\partial_j\chi^{(1)}\partial_j e^{(2)}-\partial_i\partial_je^{(2)}\partial_j \chi^{(1)}\right] \, , \label{j}
\end{eqnarray}
where in the last expression $e$ is computed from \eqref{e} and, to lowest order, from \eqref{cr2}. This fixes step 1.

For steps 2 and 3 we consider the field equations given by \eqref{field_eq} and \eqref{khronon_eq}.

We begin the computation with the khronon equation (\ref{khronon_eq}). To lowest order, using \eqref{j} we obtain 
\begin{equation}
-\Delta\chi^{(1)}\Delta e^{(2)}+2\del_i\del_j\chi^{(1)}\del_i\del_je^{(2)}+\del_i\chi^{(1)}\del_i\Delta e^{(2)}=\Delta \dot{e}^{(2)}\,.
\end{equation}
This is solved if we require $e^{(2)}=0$. From (\ref{e}) this is equivalent to $\mathcal{R}^{(2)}=0$ which, from (\ref{cr2}), determines $\chi^{(1)}$ through
\begin{equation}
-2\Delta\chi^{(1)}\Delta\chi^{(1)}+2\Delta\dot{\chi}^{(1)}-2\partial_i\chi^{(1)}\partial_i\Delta \chi^{(1)} =-\Delta h_{00}^{(2)}+\delta_{ij}\Delta h_{ij}^{(2)}-\partial_{ij}h_{ij}^{(2)}
\,.\label{khronon1}
\end{equation}
If this last condition holds, the $e$ and $\varphi$ energy-momentum tensors (\ref{T_e})-(\ref{T_phi}) therefore vanish to $\mathcal{O}(2)$, and the field equations (\ref{field_eq}) reduce to those of GR (\ref{field_eq_gr}). Then, with solutions \eqref{newper} and \eqref{ppngr42} the khronon perturbation equation (\ref{khronon1}) becomes
\begin{equation}
\Delta\dot{\chi}^{(1)}-\Delta\chi^{(1)}\Delta\chi^{(1)}-\partial_i\chi^{(1)}\partial_i\Delta \chi^{(1)} =\Delta U=-4\pi\rho\,.\label{chi1}
\end{equation}
Thus, determining the first order perturbation of the khronon field by the matter density in this way, guarantees degeneracy with GR to this order. Although an analytical solution of equation (\ref{chi1}) for general $\rho$ is not available, let us note as an example that for a point-like source at rest, 
\begin{equation}
\rho=m\,\delta(\vec{r})\,,\qquad U=\frac{m}{r}\,,
\end{equation}
one can ignore the time derivative and equation (\ref{chi1}) becomes 
\begin{equation}
\nabla^i\left(\del_i\chi^{(1)}\Delta\chi^{(1)}-\frac{m}{r^2}\right)=0\,,
\end{equation}
whose simplest solution is
\begin{equation}
\chi^{(1)}=\pm\sqrt{\frac{8}{3}mr}\,.\label{chi1_sol}
\end{equation}
Note that although $\chi^{(1)}$ grows with $r$, the physical meaning is in its derivative, which conveniently decays as $1/\sqrt{r}$.\footnote{Observe also that more involved solutions for $\chi^{(1)}$ exist, but they do not change the main feature, which is that the first order PPN expansion of $n$-DBI gravity reduces to that of GR.}

The pattern should now be clear. Inspection of the khronon equation (\ref{khronon_eq}) reveals that $e={\rm constant}$ is a solution. From (\ref{e}) this is equivalent to $\mathcal{R}=0$ to the appropriate order. Demanding $\mathcal{R}=0$ to a given order, the field equations (\ref{field_eq}) reduce to those of GR (\ref{field_eq_gr}) and so does the PPN or the post-PPN solutions. Observe that whereas $\chi^{(1)}$ will only depend on $\rho$, the next order perturbation, $\chi^{(3)}$, will depend also on $v^i$, $\Pi$ and $p$; in fact a computation of the condition $\mathcal{R}^{(4)}=0$ determines that $\chi^{(3)}$ obeys
\begin{equation}
\Delta\dot{\chi}^{(3)}-2\Delta\chi^{(1)}\Delta\chi^{(3)}-\del^i\chi^{(1)}\del_i\Delta\chi^{(3)}-\del^i\chi^{(3)}\del_i\Delta\chi^{(1)}=\frac{1}{2}R^{(4)}+(\dots) \ ,\label{chi3}
\end{equation}
where $(\ldots)$ does not contain $\chi^{(3)}$, i.e. is already determined by the metric and $\chi^{(1)}$. The structure of the equations determining $\chi^{(1)}$ and $\chi^{(3)}$ is therefore quite similar (actually, equation (\ref{chi3}) is simpler than (\ref{chi1}) since it is linear in $\chi^{(3)}$).

For the solution (\ref{chi1_sol}), equation (\ref{chi3}) becomes a second order ODE for $\chi'^{(3)}(r)$ whose solution is
\begin{equation}
\chi'^{(3)}(r)=\pm\frac{1}{\sqrt{6mr}}\int_1^r dx\, RHS(x)\left(\frac{x^5}{r^3}-x^2\right)\,,
\end{equation}
where $RHS(x)$ is the right-hand side of equation (\ref{chi3}). One can check that the $r$-dependence of the terms in $(\ldots)$ ensure that $\chi'^{(3)}(r)$ doesn't grow with $r$. For example, $\Delta\chi^{(1)}\del^i\chi^{(1)}\del_i\chi^{(1)}$ and $h_{ij}^{(2)}\del_i\del_j\chi^{(1)}$ both go as $r^{-5/2}$.

\subsection{Discussion}
\label{discussion_PPN}
Contrary to our expectations, the solution just found is nothing but the usual GR solution. Our derivation relied on the ability to set $\mathcal{R}=0$, which is a sufficient condition for a GR solution to be also a solution of $n$-DBI  \cite{Herdeiro:2011im,Coelho:2013zq}. From (\ref{R_def}), this means solving
\begin{equation}
\nabla_\mu(n^\mu\nabla_\nu n^\nu)=-\frac{1}{2}R\,,\qquad n_\mu=-\frac{\del_\mu\varphi}{\sqrt{-g^{\mu\nu}\del_\mu\varphi\del_\nu\varphi}}\,,
\end{equation}
for $\varphi$. This \emph{slicing condition} is certainly not feasible for generic space-times, but it seems likely that a perturbative expansion around a background solution can be found order by order. Most importantly, setting $\mathcal{R}=0$ does not impose any additional constraints on the two parameters of $\lambda,q$. This subset of solutions exists for any $\lambda,q$, and therefore this analysis leaves the parameters of $n$-DBI gravity unconstrained apart from the requirement that the effective cosmological constant is small (which we assumed when choosing Minkowski as the background solution).  In particular, we conclude that $n$-DBI gravity predicts no post-Newtonian preferred-frame effects, a counter-intuitive property given its similarity to Ho\v rava-type theories. In fact, at first PPN order, $n$-DBI gravity is indistinguishable from GR and it is plausible this equivalence remains at higher orders.


\section{Scalar Perturbations of FRW in $n$-DBI}
\label{scalarfrw}
The results of Section \ref{ppn} show that $n$-DBI is consistent with solar system experiments and provide an indication that the theory matches the predictions of GR at higher perturbation theory (PN) order around flat space-time. An experimental signature of this theory, therefore, must be searched for in a different arena. In this section we will discuss gravitational scalar perturbations around a homogeneous and isotropic (FRW) cosmological background. Our ultimate goal is to analyze if the perturbations of the scalar graviton may give rise to a power spectrum compatible with observational constraints.

\subsection{Scalar perturbations of a spatially flat FRW model}
We want to study gravitational scalar perturbations around a spatially flat FRW model in $n$-DBI gravity. Using a conformal time coordinate, $\tau$, and denoting by $\phi(\tau)$ the conformal factor, the geometry reads
\begin{equation}
ds^2=\phi^2(\tau)\left[-(1+2A)d\tau^2+2\del_i Bdx^id\tau+\left((1-2\psi)\delta_{ij}-2\del_i\del_jE\right)dx^i dx^j\right]
\,,
\label{pertfrw}
\end{equation}
The short-hands $\mathcal{H}\equiv\dot{\phi}/\phi$ and $\kappa\equiv G_N/6\lambda$ will be used, and the Laplace operator $\Delta$ shall be interpreted as acting in Fourier space $(\equiv-k^2)$ where needed.
As in Section \ref{ppn}, the khronon field will also be perturbed as
\begin{equation}
\varphi=\tau+\chi\,.
\end{equation}
Under a gauge transformation
\begin{equation}
\tau\rightarrow\tau+T\,,\qquad x^i\rightarrow x^i+\del^iL\,,
\end{equation}
these perturbations transform as
\begin{eqnarray}
\chi&\rightarrow&\chi+T\,,\\
A&\rightarrow&A+\dot{T}+\mathcal{H}T\,,\\
\psi&\rightarrow&\psi-\mathcal{H}T\,,\\
B&\rightarrow&B-T+\dot{L}\,,\\
E&\rightarrow&E-L\,.
\end{eqnarray}

\subsection{Flat background}
Before going to the cosmological relevant de Sitter background, we shall review the behavior of the scalar mode around flat Minkowski space-time \cite{Coelho:2012xi}. This is given by $\phi(\tau)=1$ in the previous expressions and $q=1$ in the action. Note that the combinations
\begin{equation}
\psi\,,\qquad A+\dot{B}+\ddot{E}
\end{equation}
are fully gauge-invariant.
The second-order Lagrangian, in unitary gauge ($\chi=0$), reads
\begin{equation}
\mathcal{L}^{(2)}=-6\dot{\psi}^2+4\psi(\Delta A+\Delta \dot{B}+\Delta\ddot{E})-2\psi\Delta\psi-\frac{1}{2}\kappa\left(\mathcal{R}^{(1)}\right)^2,
\end{equation}
where
\begin{equation}
\mathcal{R}^{(1)}=4\Delta\psi-2\Delta A\,.
\end{equation}

The three independent equations of motion can be cast as
\begin{eqnarray}
\Delta\dot{\psi}&=&0\,,\label{flat1}\\
\Delta\psi&=&-\frac{\kappa}{4}\Delta\mathcal{R}^{(1)}\,,\label{flat2}\\
\Delta(\dot{B}+\ddot{E}+A+\psi)&=&0\label{flat3}\,,
\end{eqnarray}
emphasizing the fact that they reduce to GR whenever $\mathcal{R}^{(1)}=0$ (or in the GR limit $\kappa\rightarrow0$). Note this is a gauge-invariant statement, even if we make use of the khronon field, in which case
\begin{equation}
\mathcal{R}^{(1)}\rightarrow4\Delta\psi-2\Delta A+2\Delta\dot{\chi}\,,\label{chi_curly_R}
\end{equation}
becomes gauge-invariant as well. It shows, however, that GR solutions are a subset of all possible solutions. In the Newtonian (or longitudinal) gauge $B=E=0$, the metric contains one arbitrary function of space,
\begin{eqnarray}
\psi(x)=-A(x)\,,\label{psiArelation}
\end{eqnarray}
which accounts for half degree of freedom, whereas the other half is hidden in the khronon perturbation and is obtained by solving (\ref{flat2}) with the substitution of (\ref{chi_curly_R}): 
\begin{eqnarray}
\chi(x,\tau)=C(x)-\left(3+{2\over\kappa\Delta}\right)\psi(x)\tau\,,\label{Khrononpertflat}
\end{eqnarray}
where $C(x)$ is an arbitrary function of space.

\subsection{de Sitter background}
The de Sitter universe is an exact solution of $n$-DBI gravity \cite{Herdeiro:2011km,Herdeiro:2011im}. Moreover, it describes the late time behaviour of the cosmological solution described in \cite{Herdeiro:2011km}. It is described by \eqref{pertfrw} without the perturbations and
\begin{equation}
\phi(\tau)=-\sqrt{\frac{3}{\Lambda}}\frac{1}{\tau}\,,\qquad \Lambda=\frac{3\lambda(1-q^{-2})}{G_N}=\frac{1-q^{-2}}{2\kappa}\,, \qquad \mathcal{H}=-\frac{1}{\tau}\,.
\end{equation}

We shall now consider the behaviour of the gravitational scalar perturbations in this particular background of $n$-DBI gravity. We recall that in standard inflationary theory, the power spectrum for density perturbations, generated by quantum fluctuations of the scalar field during the inflationary era, is obtained by studying a canonical scalar field on a de Sitter background in GR. Thus our study aims at testing how the perturbations of the $n$-DBI gravity scalar mode compare with the behaviour of scalar perturbations in the standard theory. 

It is convenient to define
\begin{equation}
\Phi_1\equiv\dot{\psi}+\mathcal{H}A\,,\qquad \Phi_2\equiv\psi-\mathcal{H}(B+\dot{E})\,,\qquad \Phi_3\equiv4\psi-A\,.
\end{equation}
$\Phi_1$ and $\Phi_2$ are fully gauge-invariant, whereas $\Delta\Phi_3$ is invariant under FPD's only. As before, however, if we introduce the khronon field,
\begin{equation}
\Phi_3\rightarrow 4\psi-A+\dot{\chi}+5\mathcal{H}\chi\,,\label{Phi3chi}
\end{equation}
then $\Phi_3$ becomes fully gauge-invariant as well.
Another set of gauge-invariants, called \emph{Bardeen potentials} \cite{Bardeen:1980kt}, are common in cosmology:
\begin{eqnarray}
\Phi_\mathcal{B}&\equiv&A+\mathcal{H}(B+\dot{E})+(\dot{B}+\ddot{E})=\mathcal{H}^{-1}(\Phi_1-\dot{\Phi}_2)\,,\\
\Psi_\mathcal{B}&\equiv&\psi-\mathcal{H}(B+\dot{E})=\Phi_2\,.
\end{eqnarray}
It turns out that the equations of motion assume a very simple form when written in terms of $\Phi_{1,2,3}$. The Hamiltonian and momentum constraints, respectively,
\begin{eqnarray}
-6\mathcal{H}(\phi^2-\kappa\Delta)\Phi_1+2(\phi^2-\kappa\Delta)\Delta\Phi_2+\kappa(\Delta-6\mathcal{H}^2)\Delta\Phi_3&=&0\,,\\
\phi^2\Delta\Phi_1-2\kappa\mathcal{H}\Delta^2\Phi_2+\kappa\mathcal{H}\Delta^2\Phi_3&=&0\,,
\end{eqnarray}
can be readily solved for $\Phi_{1,2}(\Phi_3)$,
\begin{equation}
\Phi_1=\frac{\kappa\Delta}{\kappa\Delta-\phi^2}\mathcal{H}\Phi_3\,,\qquad \Phi_2=\frac{1}{2}\frac{\kappa\Delta}{\kappa\Delta-\phi^2}\Phi_3\,.\label{Phi1Phi2}
\end{equation}
The evolution equation,
\begin{equation}
3\dot{\Phi}_1+6\mathcal{H}\Phi_1-\Delta\Phi_2+\mathcal{H}^{-1}\Delta(\Phi_1-\dot{\Phi}_2)=\kappa\phi^{-2}\left(12\mathcal{H}\Delta\Phi_1-2\Delta^2\Phi_2+(\Delta-3\mathcal{H}^2)\Delta\Phi_3-3\mathcal{H}\Delta\dot{\Phi}_3\right)\,,
\end{equation}
upon using (\ref{Phi1Phi2}), simplifies to a first-order in time equation for $\Phi_3$,
\begin{equation}
\dot{\Phi}_3-\mathcal{H}\frac{5\phi^2-3\kappa\Delta}{\phi^2-\kappa\Delta}\Phi_3=0\,.
\end{equation}
The solution, with an arbitrary function of wavenumber $k$, $C_1(k)$, can be written as
\begin{eqnarray}
\Phi_{1,k}(\tau)&=&2\sqrt{\Lambda\over 3}C_1(k)\phi(\tau)^4\,,\qquad \Phi_{2,k}(\tau)=C_1(k)\phi(\tau)^3\,,\\
\Phi_{3,k}(\tau)&=&2C_1(k)\phi(\tau)^3\left(1+\frac{\phi(\tau)^2}{\kappa k^2}\right)\,.\label{solPhi123}
\end{eqnarray}
Then the Bardeen potentials,
\begin{equation}
\Psi_{\mathcal{B}}=-\Phi_{\mathcal{B}}=\Phi_2\,,
\end{equation}
are enough to completely determine the metric in the conformal Newtonian (or longitudinal) gauge $B=E=0$. The function $C_1(k)$ accounts for half degree of freedom. The other half appears when solving (\ref{solPhi123})  for $\chi$ by substituting (\ref{Phi3chi}),
\begin{equation}
\chi_k(\tau)=-\sqrt{3\over\Lambda}C_1(k)\left({3\phi(\tau)^2\over 7}-{2\phi(\tau)^4\over 9\kappa k^2}\right)+{C_2(k)\over \phi(\tau)^5}\,.\label{Khrononpert}
\end{equation}

The scalar graviton of $n$-DBI gravity grows as the universe expands, i.e., as $\phi(\tau)$ increases.
Firstly, this contrasts with a canonical scalar field in General Relativity, which oscillates on subhorizon scales ($-k\tau>>1$) but becomes frozen once it crosses the horizon (see \cite{Riotto:2002yw,Langlois:2010xc} for pedagogical reviews; and \cite{Gao:2009ht} for cosmological perturbations in Ho\v{r}ava-Lifshitz gravity). 
Secondly, this result indicates that de Sitter space is unstable in $n$-DBI gravity, and the lifetime of de Sitter universe is set by the cosmological constant, $T_{\rm life}\sim 1/\sqrt{\Lambda}$.

\subsubsection{Flat limit $\Lambda\rightarrow0$}
The coordinates $(\tau,x^i)$ that we have been using are ill-defined in the limit $\Lambda\rightarrow0$. Therefore, we introduce the new time coordinate $T$ defined by
\begin{equation}
\tau=-\sqrt{3\over\Lambda}e^{-\sqrt{\frac{\Lambda}{3}}T}\,,
\end{equation}
or equivalently $\phi(\tau)=e^{\sqrt{\frac{\Lambda}{3}}T}$.
Then the metric ansatz (\ref{pertfrw}) becomes
\begin{equation}
ds^2=-(1+2A)dT^2+2\del_iBdx^i e^{\sqrt{\frac{\Lambda}{3}}T}dT+e^{2\sqrt{\frac{\Lambda}{3}}T}\left((1-2\psi)\delta_{ij}-2\del_ i\del_jE\right)dx^idx^j\,.\label{metric2}
\end{equation}
The Big Bang was at $\tau=-\infty$ and is now at $T=-\infty$, but the distant future $\tau=0^-$ has been mapped to $T=+\infty$. The limit $\Lambda\rightarrow0$ is now well defined. It is useful to note that, in this limit,
\begin{equation}
\phi\rightarrow 1\,,\qquad \tau\rightarrow-\sqrt{3\over \Lambda}\,,\qquad \mathcal{H}\rightarrow \sqrt{\Lambda\over 3}\,.
\end{equation}
The gauge invariant perturbation is now
\begin{eqnarray}
\Phi_{1,k}(T)&=&2\sqrt{\Lambda\over 3}C_1(k)e^{4\sqrt{\frac{\Lambda}{3}}T}\,,\qquad \Phi_{2,k}(T)=C_1(k)e^{3\sqrt{\frac{\Lambda}{3}}T}\,,\\
\Phi_{3,k}(T)&=&2C_1(k)e^{3\sqrt{\frac{\Lambda}{3}}T}\left(1+\frac{e^{2\sqrt{\frac{\Lambda}{3}}T}}{\kappa k^2}\right)\,.\label{solPhi123flat}
\end{eqnarray}
In the limit $\Lambda\rightarrow0$, it can be easily checked that $\psi_k=-A_k=C_1(k)$ in the Newtonian gauge in consistent with (\ref{psiArelation}). It is also straightforward to show that the khronon perturbation (\ref{Khrononpert}) reduces to (\ref{Khrononpertflat}).\footnote{Care is needed when taking the $\Lambda\to 0$ limit. As it turns out, it is most appropriate to consider $\phi^5\chi_k$ instead of $\chi_k$. The time independent divergent piece proportional to $\sqrt{3/\Lambda}C_1(k)$ can be absorbed by a trivial redefinition of $C_2(k)$.}


\section{Discussion and final remarks}
\label{conclusion}
GR has successfully passed a battery of observational tests in astrophysical systems but it is in tension with cosmology. The GR based FRW cosmological model can only be compatible with observations by invoking two dark components, plus adding an extra ingredient - typically in the form of one or many scalar fields - that sources an early inflationary epoch, widely accepted to be required for consistency of the picture. All these ingredients are, at present, mysterious at the level of fundamental physics, and as such, exotic. Thus, it is important to explore alternative models of gravity that may improve the cosmological picture - by requiring less exotic ingredients - but simultaneously keeping the astrophysical predictions of GR. 

$n$-DBI gravity provides an example where these two aims are achieved in a novel way.  On the one hand, solar system predictions of GR will be kept, as shown in the first part of this paper, because the theory contains, as a subsector, a class of solutions that matches solutions of GR. And this does not hold only for a discrete set of exact solutions as shown before \cite{Herdeiro:2011im,Coelho:2013zq}; this subsector has a self-contained perturbation theory, so that the PPN parameterization coincides, at first order, with that of GR. In a sense, this subsector is a \textit{consistent truncation of $n$-DBI gravity}. Moreover, this subsector does not depend on the two adjustable parameters of $n$-DBI gravity and hence solar system tests do not constrain them. As we have argued at the end of Section \ref{discussion_PPN}, it seems likely that by adjusting the khronon field that controls the space-time foliation, the matching with the post-Newtonian expansion of GR will hold at higher orders as well and hence that $n$-DBI will equally pass all astrophysical tests provided by compact binary systems. The analysis performed in Sec. \ref{ppn}, however, was not exhaustive, in the sense that we have not shown that the solution provided, which matches that of GR, is unique. This is an important open question. 

On the other hand, the cosmological solution that matches observations \cite{Herdeiro:2011km} is not in this subsector; in other words, it is not a solution of GR (with the same energy momentum tensor). This suggests we might find experimental signatures of $n$-DBI by analyzing cosmological perturbations. As a first step towards that goal we have considered in Section \ref{scalarfrw} generic gravitational scalar perturbations in $n$-DBI gravity.
Specializing to de Sitter space-time we have found solutions for this scalar perturbation which grow with time as the universe expands. Moreover, performing a mode decomposition, we observe that there is no freezing of the oscillations on super-horizon scales, in contrast with the behaviour of scalar fields in de Sitter, in GR. This provides an example on how cosmological perturbations computed in a model where Lorentz invariance is broken may differ, conceptually, from the standard GR case. 
Although it is not clear if and to what extent this qualitative feature applies to scalar perturbations in the inflationary era of the cosmological solution in \cite{Herdeiro:2011km}, the detailed computation of the power spectrum of scalar perturbations will presumably have to take into account the detailed evolution of the universe, from the inflationary to the non-accelerating epochs, in order to obtain the amplitude of the different modes as they re-enter the horizon. 
We expect to report on the power spectrum obtained in this cosmological solution in the near future. 

Finally, let us comment on the Parameterized Post-Friedmannian (PPF) formalism \cite{Baker:2012zs}, developed to compare the cosmological predictions of a large set of modified gravity theories in a unified framework, much in the same way as the PPN formalism is used to compare deviations from Newtonian gravity for different theories of gravity. The PPF formalism gives an ansatz for the equations of motion, which are required to be covariant and at most second order in time. Although the ADM formulation of $n$-DBI \cite{Herdeiro:2011im}, as defined by the Hamiltonian constraint, momentum constraint and evolution equation for the $3$-metric, contains no more than two time derivatives, its covariant form, as presented in Section \ref{intro_ppn}, necessarily contains higher time derivatives in the equation of the khronon (St\"uckelberg) field. Indeed,  even in unitary gauge, the extra equation for the scalar mode, obtained by taking the time derivative of the Hamiltonian constraint, is higher order in time \cite{Coelho:2012xi}. This makes the PPF inapplicable to $n$-DBI gravity, at least in this formulation. Curiously, the PPF framework can be applied to Ho\v{r}ava-Lifshitz gravity. The reason why it works for Ho\v{r}ava-Lifshitz gravity can be traced to the the fact that the action contains only one derivative of the vector field $n^\mu$, whereas $n$-DBI has two derivatives. As shown in \cite{Blas:2009yd}, this makes the linearised equation for the scalar mode, in unitary gauge, only second-order in time, despite being generically higher-order.

\section*{Acknowledgements}
F.C. and C.H. are very grateful to the physics department of Nagoya University for hospitality, where part of this work was done. 
S.H. would like to thank the Yukawa Institute for Theoretical Physics for their hospitality.
Y.S. thanks National Center for Theoretical Sciences (North), National Taiwan University, Tunghai University and the Yukawa Institute for Theoretical Physics for comfortable hospitality. Y.S. would like to appreciate the Hakubi Center for Advanced Research for financial support. 
This work was supported by the {\it NRHEP--295189} FP7-PEOPLE-2011-IRSES Grant, by
  FCT -- Portugal through the project PTDC/FIS/116625/2010, by FCT strategic project PEst-C/CTM/LA0025/2011 and by Funda\c c\~ao Calouste Gulbenkian through the ``Programa de Est\'\i mulo \`a Investiga\c c\~ao''. The work of S.H. was supported in part by the Grant-in-Aid for Nagoya University Global COE Program (G07) and the South African Research Chairs Initiative of the Department of Science and Technology and National Research Foundation.


\appendix
\renewcommand{\theequation}{\Alph{section}.\arabic{equation}}


\end{document}